\documentclass[11pt]{article}
\usepackage{epsfig}
\usepackage{amsthm,amsmath}
\numberwithin{equation}{section}

\newcommand{\Dfig}[2]{\epsfig{figure=#1.eps,width=#2cm}}

\newenvironment{Eonefigs}[3]
{
        \begin{figure} [t]
          \begin{center}
          \begin{tabular}{c}
              \Dfig{#1}{#3} \\
          \vspace{-.5in}
          \end{tabular}
          \caption{{\footnotesize #2}}
          \label{#1}
          \end{center}
        \end{figure}
}{}

\newenvironment{Etwofigs}[4]
{
        \begin{figure} [h]
          \begin{center}
          \begin{tabular}{cc}
          \Dfig{#1}{#4}&\Dfig{#2}{#4} \\
          \vspace{-.25in}
          \end{tabular}
          \caption{\footnotesize #3}
          \label{#1}\label{#2}
          \end{center}
        \end{figure}
}{}

\begin{document}

\title{The Inverse Simpson Paradox\\
(How to win without overtly cheating)}

\author{Ora E. Percus and Jerome K. Percus\\
Courant Institute of Mathematical Sciences\\
New York University\\
251 Mercer Street\\
New York, NY  10012\\
Email:percus@cims.nyu.edu}

\date{\today}
\maketitle

\begin{abstract}
Given two sets of data which lead to a similar statistical
conclusion, the Simpson Paradox [10] describes the tactic of
combining these two sets and achieving the opposite conclusion.
Depending upon the given data, this may or may not succeed.  Inverse
Simpson is a method of decomposing a given set of comparison data
into two disjoint sets and achieving the opposite conclusion for
each one. This is always possible; however, the statistical
significance of the conclusions does depend upon the details of the
given data.
\end{abstract}

\section{Introduction}
Anyone contemplating a statistical analysis is warned, at an early
stage of the game, ``but don't combine the statistics of monkey
wrenches and watermelons'', or the equivalent.  Failure to heed this
instruction -- at a more sophisticated level, to be sure -- gives
rise frequently to Simpson's Paradox (here, in its 2-trial sequence
version): if choice $A$ is ``statistically better'' than choice $B$
in each of two sets of trials under differing circumstances, then it
may happen that merging the two sets of data produces the opposite
conclusion.  Consider the following specially constructed example
for the sake of illustration:
\begin{table}[htbp]
\caption{Simpson Paradox Prototype}\label{Ta:2}
\begin{center}
\begin{tabular}{l c c c}
&Trial \#1 &Trial \#2 &Total\\
$S_A\equiv$ $A$ successes &60 &60& 120\\
$F_A\equiv$ $A$ failures &20 &140 &160\\
$S_B\equiv$ $B$ successes &140 &20 &160\\
$F_B\equiv$ $B$ failures &60 &60 &120\\
\multicolumn{4}{c}{60/80$>$140/200 and 60/200$>$20/80}\\
\multicolumn{4}{c}{but 120/280$<$160/280}\\
\end{tabular}
\end{center}
\end{table}

In Fig. 1, we pictorially represent trial sequence \#1 by a solid
line, trial sequence \#2 by a dashed line; trial \#1 tests drug $A$,
$N_1$ times, drug $B$, $N_2$ times, while trial \#2 reverses the
number of tests. The successes $S$, and failures $F$ are shown for
each drug in each trial sequence.  If $a<b$, so $1-a>1-b$, then
clearly the $S/N$ ratio of drug $A$ is larger than that of $B$ in
both trial sequences, so drug $A$ certainly seems better.  But in
the combined trials $S_A/N_A=((1-a)N_1+bN_2)/(N_1+N_2)$ is lower
than $S_B/N_B=((1-b)N_2+aN_1)/(N_1+N_2)$ if $(1-a)
N_1+bN_2<(1-b)N_2+aN_1$, or
\begin{equation}
N_1<\frac{1-2b}{1-2a}\,N_2,
\end{equation}
a quite feasible circumstance, so that drug $A$ has now become
inferior to $B$!

\Etwofigs{Fig1}{Fig2}{Simpson Paradox Prototype}{3.7}

This phenomenon is well-known and well-documented [5] [6] [7] [8]
[9] [10] -- but hope springs eternal.  Only recently [1], a drug
manufacturer, whose current potential blockbuster drug (Xinlay)
failed to better a placebo in two clinical trials with uncorrelated
protocols, proposed to a regulatory agency to pool the two
sequences.  If accepted, their drug would then outperform the
placebo, allowing them to move forward.  The regulatory agency panel
was not unaware of the forced paradox, and denied the
reinterpretation of the data.

\section{Inverse Simpson}

The Simpson Paradox is data-driven.  As in (1.1), it may, or may
not, hold in a given situation.  However, what we may term inverse
Simpson paradox is a different story: can we take a long pair of
data streams -- say successes and failures with drug $A$, and
similarly with drug $B$ -- and decompose them into two pairs of
subsequences, each of which reverses the conclusion of the original
pair?  This can be carried out in different ways and for different
purposes,
\begin{enumerate}
\item[a)] Most directly and legitimately, it may be realized that data
from two sources were combined for simplicity, and so there is a
unique decomposition called for, which may indeed reverse the
conclusion.  This appears to be the case in the oft-quoted Berkeley
sex discrimination controversy [5].
\item[b)] Least directly and least legitimately -- but perhaps an effective
strategy in litigation -- one can ask for that decomposition that
maximally reverses the conclusion, and then use ingenuity to
characterize the subsets thus obtained.
\item[c)] Putting a different spin on b), one can ask for that decomposition
that maximally comes jointly to either conclusion, and use this as
an investigative tool to recognize a hidden characterization of
significant subsets of related entities.
\end{enumerate}

At first blush, inverse Simpson, in contexts b) and c), is trivially
accomplished.  Fig. 2 illustrates the principle.

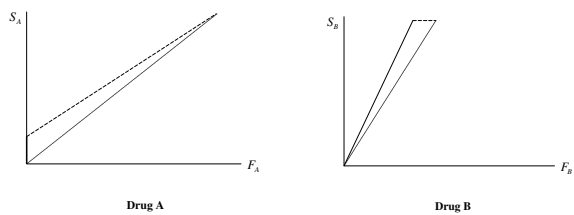
\Etwofigs{Fig3}{Fig4}{Inverse Simpson Prototype}{3.7}

The dotted lines refer to the assertedly pooled data, clearly
indicating that $A$ loses to $B$.  The hypothetical trial 1 data is
represented by solid lines, and since $A$ has only successes, it is
surely superior.  And the dashed lines refer to trial 2, in which
$B$ has only failures, and so surely loses.

But Fig. 2 is a suspiciously extreme version of a strategy that can
be made to look more reasonable.  To put it in context, let us
consider the well-known Berkeley sex discrimination case [5], which
we will paraphrase for numerical simplicity.  The original data is
that in one division, $S_A=41$ out of $N_A=100$ male applicants were
admitted, a success rate of $P_A=.41$.  On the other hand, $S_B=29$
of $N_B=100$ female applicants were admitted, a success rate of only
$P_B=.29$.  Clearly, it would seem that the admission process
discriminated against females. This was not the case. In fact,

\begin{table}[htbp]
\caption{Simplified Berkeley Admission Data}\label{Ta:2}
\begin{center}
\begin{tabular}{l c c }
&Dept. 1 &Dept. 2\\
Male Applicants &30 &70\\
Males Admitted &6 &35\\
Female Applicants &70 &30\\
Females Admitted &14 &15\\
\multicolumn{3}{r}{Total Male Admissions/Applicants 41/100=.41}\\
\multicolumn{3}{r}{Total Female Admissions/Applicants 29/100=.29}\\
\end{tabular}
\end{center}
\end{table}

\noindent Table 2, Simplified Berkeley Admission Data, was arrived
at by combining that of two departments, say 1 and 2. Referring to
Table 2, we see that the success rates of males in the two
departments were $P_{A1}=.2$, $P_{A2}=.5$, with the corresponding
female success rates of $P_{B1}=.2$, $P_{B2}=.5$. There was no
demonstrable discrimination in either department, but ``mixing
watermelons and monkey wrenches'' created very much of a statistical
artifact.

Let us proceed to a general situation.  We are given $N_A$ and
$P_A=S_A/N_A,\, N_B$, and $P_B=S_B/N_B$ for which, without loss of
generality, $P_A>P_B$.  We then imagine compartmentalizing the
$A$-pool as $N_{A1}=\alpha N_A$, $N_{A2}=(1-\alpha)N_A$, and the
$B$-pool as $N_{B1}=\beta N_B$, $N_{B2}=(1-\beta)N_B$; the success
rates are to be given via $S_{A1}=P_{A1}N_{A1}$,
$S_{A2}=P_{A2}N_{A2}$, $S_{B1}=P_{B1}N_{B1}$, $S_{B2}=P_{B2}N_{B2}$.
The question then is whether $\alpha$ and $\beta$ can be chosen so
that
\begin{equation}
\begin{aligned}
P_{A1}&=\lambda=P_{B1}\\
P_{A2}&=\mu=P_{B2},
\end{aligned}
\end{equation}
indicating no advantage to $A$ or $B$ in either case.  This is
trivial.  Since $S_{A1}=\alpha\lambda N_A$, $S_{A2}=(1-\alpha)\mu
N_A$, $S_{B1}=\beta\lambda N_B$, $S_{B2}=(1-\beta)\mu N_B$, we must
have
\begin{equation}
\begin{aligned}
P_A&=\alpha\lambda+(1-\alpha)\mu\\
P_B&=\beta\lambda+(1-\beta)\mu
\end{aligned}
\end{equation}

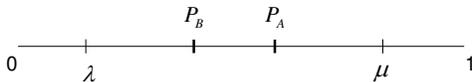
\Eonefigs{Fig5}{Placement of Averaging Parameters $\lambda$ and
$\mu$}{7}

Thus, $P_A$ and $P_B$ are both averages of $\lambda$ and $\mu$,
which therefore must lie outside the interval $(P_B, \,P_A)$ as in
Fig. 3.  Explicitly, of course, we have
\begin{equation}
\begin{aligned}
\alpha&=\frac{\mu-P_A}{\mu-\lambda}
&\beta&=\frac{\mu-P_B}{\mu-\lambda}\\
1-\alpha&=\frac{P_A-\lambda}{\mu-\lambda}
&1-\beta&=\frac{P_B-\lambda}{\mu-\lambda}
\end{aligned}
\end{equation}
In situations not as clear cut as the Berkeley case, we would want
to invent a hypothetical decomposition in which e.g. $\lambda$ is
roughly in the middle of the $(0,\,P_B)$ interval, $\mu$ roughly in
the middle of $(P_A,1)$, in order to allay suspicion.  In the
Berkeley case, we see that $\lambda = .2$, $\mu=.5$ do satisfy this
criterion.

With (2.3), we find that a suitable decomposition removes the
apparent bias against females: no assertion can then be made.  But
Fig. 2 illustrates a proactive strategy, in which a suitable
decomposition reverses the original assertion and appears to
establish the superiority of $A$.  What is wrong with the
construction of Fig. 2, aside from its suspicious extreme nature?
Nothing, but the conclusion is questionable because we have not
attended to the statistical significance of the new assertions, a
point that was emphasized by the FDA panel cited above.  Doing so
forms the substance of our ensuing discussion.

\section{Statistical Significance}

A prototypical situation calling for statistical assessment is this.
A sequence of $N$ independent Bernoulli trials -- successes or
failures -- is carried out on the same object, resulting in $S$
successes.  Given $\epsilon$, with what probability, or
\underline{confidence}, can we claim that $p$, the intrinsic success
probability parameter, satisfies
\begin{equation}
|p-S/N|\le\epsilon/N^{1/2}?
\end{equation}
The standard approach is to start with the elementary result that,
regarding $S$ as a random variable and defining $q\equiv 1-p$,
\begin{equation}
\begin{aligned}
Pr(&|S-Np|\le N^{1/2}\epsilon|p)\\
&=\sum_{j=[Np-N^{1/2}\epsilon]}^{[Np+N^{1/2}\epsilon]}
\left(\begin{smallmatrix}{N}\\
{j}\end{smallmatrix}\right)p^jq^{N-j},
\end{aligned}
\end{equation}
where [ ] denotes integer part.  The device then is to identify
(3.2), which is a probability on $S$-space, with a probability on
$p$-space:
\begin{equation}
Pr(|p-S/N|\le\epsilon/N^{1/2}|S)=Pr(|S-Np|\le N^{1/2}\epsilon|p)
\end{equation}
signifying our confidence that (3.1) holds.

The sort of information that will interest us will, however, in the
context of this prototype, be more like: with what confidence, based
upon the observed value of $S$, can we claim that
\begin{equation}
p\ge 1/2?
\end{equation}
Now, the above recipe is not readily applicable, since we are no
longer questioning a \underline{relationship} between $p$ and $S$
that makes possible the sub rosa journey from $S$-space to
$p$-space. But this is indeed the province of the Bayes approach [4]
which -- ignoring the controversy that continues to swirl around it
-- is what we will use. First of all, let up recall what (3.1) would
become in a Bayesian context: we imagine joint $(p,S)$-space and
quote the obvious
\begin{equation}
\begin{aligned}
Pr(p=p'|S=S')&=Pr(S=S'|p=p')\,f(p')/Z\\
\text{where}\,Z&=\int'_0 Pr(S=S'|p=p'')\,f(p'')dp'',
\end{aligned}
\end{equation}
$f$ here referring to probability density.  If $f(p')$ is the prior
density on $p$-space, then
\begin{equation}
\begin{aligned}
Pr(|p-S/N|&\le\epsilon/N^{1/2}|S=S')\\
&=\int^{S'/N+\epsilon/N^{1/2}}_{S'/N-\epsilon/N^{1/2}}f(p')p'^{S'}q'^{N-S'}dp'/Z\\
Z&=\int'_0 f(p')p'^{S'}q'^{N-S'} dp'.
\end{aligned}
\end{equation}
But suppose we choose a uniform prior, $f(p)=1$; then (3.6) becomes
\begin{equation}
\begin{aligned}
Pr(|p-S/N|\le\epsilon/ N^{1/2})\\
=\int^{\min
(S'/N+\epsilon/N^{1/2},N)}_{\max(O,S'/N-\epsilon/N^{1/2})}p'^S q'^{N-S}dp'/Z\\
Z=\int'_0 p'^S
q'^{N-S}dp'=((N+1)\left(\begin{smallmatrix}{N}\\
{S}\end{smallmatrix}\right))^{-1}.
\end{aligned}
\end{equation}
Eqs. (3.2, 3.3) and (3.7) are certainly not identical, but if we go
to the large sample regime, i.e. the normal approximation to the
binomial, then (3.2, 3.3) aver that
\begin{equation}
Pr(|p-S/N|\le\epsilon/N^{1/2})=\int^{\epsilon/\sqrt{pq}}_{-\epsilon/\sqrt{pq}}e^{-\frac{1}{2}s'^2}ds'/\sqrt{2\pi},
\end{equation}
which, it is easy to show \underline{is} identical with the large
$N$, fixed $S/N$, steepest descent expansion [3] of (3.7) around
$p'=S/N$.

On the basis of the above equivalence, we now go immediately to the
question indicated by (3.4).  Using Bayes with a uniform prior,
precisely as in (3.7), we have
\begin{equation}
\begin{aligned}
Pr\left(p\ge\frac{1}{2}\right)=\int^1_{1/2}p'^S q'^{N-S}dp'/\int^1_0
p'^S q'^{N-S}dp'\\
=1-B_{1/2}(S+1, N+1-S)/B(S+1, N+1-S),
\end{aligned}
\end{equation}
where $B$ is the Beta function, $B_{1/2}$ the corresponding
incomplete Beta function [2].  Eq. (3.9) can also be written in the
neat form
\begin{equation}
\begin{aligned}
Pr\left(p\ge\frac{1}{2}\right)&=1-\sum_{j=0}^{N-S}\left(\begin{smallmatrix}{N+1}\\
{j}\end{smallmatrix}\right)p^{N+1-j}q^j|_{p=\frac{1}{2}}\\
&=1-\sum^{N-S}_{j=0}\left(\begin{smallmatrix}{N+1}\\
{j}\end{smallmatrix}\right)/2^{N+1}
\end{aligned}
\end{equation}
The important point however is that this construction leads quite
directly to evaluation of quantities such as $Pr(p_A\ge p_B)$, that
are appropriate to the Simpson paradox.

\section{Level of Significance of the Inverse Paradox}

The effect we are studying is not very subtle, and so it is
sufficient to take a large sample limit, which strategy we adopt.
However, there are several sample parameters, leading to the
meaningful use of additional limiting operations.  Consider first
the prototype, Eq. (3.10); here,
\begin{equation}
\alpha_N(S)=\sum_{j=0}^{N-S}\left(\begin{smallmatrix}{N+1}\\
{j}\end{smallmatrix}\right)/2^{N+1}
\end{equation}
expresses the level of significance of the assertion that
$p\ge\frac{1}{2}$, and it is not until such an assessment is made
that one can declare meaningful comparisons.  Let us evaluate (4.1)
in the large sample limit in a familiar fashion that extends at once
to the question of $Pr(p_A\ge p_B)$ relevant to the Simpson paradox.

Although (4.1) is finite and explicit, its implementation for large
$N$ and $S$ -- while trivial numerically -- is a bit complex.  For
this purpose, the expression (3.9) is more useful; it says that
\begin{equation}
\alpha_N(S)=\int^{1/2}_0 p^S(1-p)^{N-S}dp/\int^1_0 p^S(1-p)^{N-S}dp.
\end{equation}
By the large sample limit, we will mean that in which
\begin{equation}
s=N^{-1/2}\left(S-\frac{1}{2}N\right)
\end{equation}
is fixed (to within $N^{-1/2}$) as $N\to\infty$, and we then ask for
\begin{equation}
\alpha(s)=\lim_{N\to\infty}\alpha_N(S).
\end{equation}
This is obtained quite directly by a steepest descent evaluation [3]
of (4.2).  The relevant integrand is now
\begin{equation}
\begin{aligned}
I(p)\equiv p^S (1-p)^{N-S}\\
=\exp \left[\left(\frac{N}{2}+N^{1/2}s\right)\,\ln\,
p+\left(\frac{N}{2}-N^{1/2}s\right)\,\ln\,(1-p)\right],
\end{aligned}
\end{equation}
with a maximum at
\begin{equation}
p_0=\frac{1}{2}+N^{-1/2}s,
\end{equation}
and a corresponding expansion starting as
\begin{equation}
I(p)=I(p_0)\exp-\left[\frac{N}{2}(p-p_0)^2/\left(\frac{1}{4}-\frac{s^2}{N}\right)\right].
\end{equation}
Hence
\begin{equation}
\begin{aligned}
\alpha(s)&=\lim_{N\to\infty}\int^{1/2}_0
e^{-2N(p-p_0)^2}dp/\int^1_0 e^{-2N(p-p_0)^2}dp\\
&=\lim_{N\to\infty}\int^{-2s}_{-N^{1/2}-2s}e^{-x^2/2}dx/\int^{N^{1/2}-2s}_{-N^{1/2}-2s}e^{-x^2/2}dx\\
&=\int^{-2s}_{-\infty}e^{-x^2/2}dx/\int^\infty_{-\infty}e^{-x^2/2}dx,
\end{aligned}
\end{equation}
immediately recognizable in a normal distribution context.

We can then proceed to the desired evaluation of
\begin{equation}
\begin{split}
Pr(p_A\ge p_B|S_A,S_B,N_A,N_B)=\\
\sideset{}{}\iint_{\substack{1\ge p_A\ge p_B\ge 0}}
[f(p_A, p_B)\,\\
Pr(S_A, S_B,|p_A, p_B, N_A\,N_B)]dp_A\,dp_B/\\
\qquad \sideset{}{}\iint_{\substack{1\ge p_A\ge 0\\1\ge p_B\ge 0}}
[f(p_A, p_B)\\
Pr(S_A, S_B|p_A,p_B, N_A, N_B)]dp_A\,dp_B.
\end{split}
\end{equation}

This is carried out in Appendix A, where we choose Bayes with
uniform prior on $p_A,p_B$ space and process (4.9) as we did (4.2).
The result is that for large $N_A,N_B,$

\begin{equation}
\begin{split}
Pr(p_A\ge p_B)=\phi
\left(\frac{S_A}{N_A}-\frac{S_B}{N_B}\left/\left(\frac{S_A(N_A-S_A)}{N_{A^3}}\right.\right.\right.\\
\left.\left.+\frac{S_B(N_B-S_B)}{N_{B^3}}\right)^{1/2}\right)\\
\text{where}\,\phi(x)=\frac{1}{\sqrt{2\pi}}\int_{-\infty}^xe^{-\frac{1}{2}y^2}dy
\end{split}
\end{equation}

Unsurprisingly, we can obtain (4.10) as well by a version of the
probability space equivalence assertion employed in (3.3).  It is
only necessary to consider the random variable
\begin{equation}
\xi=\frac{S_A}{N_A}-\frac{S_B}{N_B}
\end{equation}
where $S_A$ and $S_B$ are binomially distributed with success
probabilities $p_A$ and $p_B$.  Since we find at once that
\begin{equation}
\begin{aligned}
E(e^{\gamma\left(\frac{S_A}{N_A}-\frac{S_B}{N_B}\right)})\\
=(p_Ae^{\gamma/N_A}+q_A)^{N_A} (p_B e^{-\gamma/N_B}+q_B)^{N_B},
\end{aligned}
\end{equation}
it follows directly that
\begin{equation}
\begin{aligned}
E(\xi|p_A,p_B)&=p_A-p_B\\
\text{Var}(\xi|p_A,p_B)&=\frac{p_A}{N_A}q_A+\frac{p_B}{N_B}q_B
\end{aligned}
\end{equation}
and then from the central limit theorem that in the limit $N_A$,
$N_B\to\infty$,
\begin{equation}
\begin{aligned}
Pr\left(\frac{S_A}{N_A}-\frac{S_B}{N_B}\ge p_A-p_B+\Delta|p_A,p_B\right)\\
=\phi(-\Delta/(p_A q_A/N_A+p_B q_B/N_B)^{1/2})
\end{aligned}
\end{equation}
The same sleight of hand as in (3.3) then converts this to
\begin{equation}
\begin{aligned}
Pr\left(p_A-p_B\le\frac{S_A}{N_A}-\frac{S_B}{N_B}-\Delta|S_A,S_B\right)\\
=\phi\left(-\Delta/\left(\frac{S_A(N_A-S_A)}{N_A^3}+\frac{S_B(N_B-S_B)}{N_B^3}\right)^{1/2}\right),
\end{aligned}
\end{equation}
and so, setting $\Delta=\frac{S_A}{N_A}-\frac{S_B}{N_B}$, to (4.10),
as was to be shown.

\section{Realizations of the Inverse Paradox}

Now let us make use of the result (4.10).  If our initial data is
characterized by $S_A$, $S_B$, $N_A+N_S=N$, and
$P_A=S_A/N_A,\,P_B=S_B/N_B$, then the confidence level with which we
can assert that $p_A\ge p_B$ is given by
\begin{equation}
\begin{aligned}
\phi&(N^{1/2}C_{AB})\\
C_{AB}&=(P_A-P_B)/\sigma_{AB}>0\\
\sigma^2_{AB}&=\frac{P_A(1-P_A)}{N_A/N}+\frac{P_B(1-P_B)}{N_B/N}.
\end{aligned}
\end{equation}
Our objective is to supply a decomposition into two hypothetical
trials $(S_{A1},N_{A1},\\S_{B1},N_{B1})$ and $(S_{A2},N_{A2},
S_{B2}, N_{B2})$ such that
\begin{equation}
\begin{aligned}
\text{if}\,C'_i&=(P_{Bi}-P_{Ai})/\sigma_i,\quad i=1,2\\
\text{where}\,P_{Ai}&=S_{Ai}/N_{Ai},\quad P_{Bi}=S_{Bi}/N_{Bi}\\
\sigma^2_i&=\frac{P_{Ai}(1-P_{Ai})}{N_{Ai}/N}+\frac{P_{Bi}(1-P_{Bi})}{N_{Bi}/N},\\
&\quad\text{then}\,C'_i>0\,\text{for}\,i=1,2.
\end{aligned}
\end{equation}
In fact, to be definite, we suppose that the two pairs of trials
reverse the initial assertion at a common level of confidence
\begin{equation}
(P_{B1}-P_{A1})/\sigma_1=C'=(P_{B2}-P_{A2})/\sigma_2
\end{equation}
with $C'>0$.  To start, we need to find the restrictions on $C'$
under which the required $(P_{A1}, P_{A2}, P_{B1}, P_{B2})$
satisfying (5.2) can be found.

The solution is direct but algebraically cumbersome, and is
presented in detail in Appendices B and C.  The conclusion of the
former is that if $\alpha\ge\beta$, then
\begin{equation}
\begin{aligned}
C'\le\min\left(\frac{\bar{\beta}\bar{P}_A-\bar{\alpha}\bar{P}_B}{\bar{\alpha}\sigma_B},
\frac{\bar{\beta}\bar{P}_A-\bar{\alpha}\bar{P}_B}{\bar{\beta}\sigma_A},\right.\\
\left.\frac{\alpha P_B-\beta P_A}{\alpha\sigma_B}, \frac{\alpha
P_B-\beta P_A}{\beta\sigma_A}\right).
\end{aligned}
\end{equation}

Since we require $C'\ge0$, this implies that
\begin{equation}
\alpha/\beta\ge
P_A/P_B\ge1,\qquad\bar{\beta}/\bar{\alpha}\ge\bar{P}_B/\bar{P}_A\ge1.
\end{equation}

In (5.4) and (5.5), we uniformly adopt the notation:
\begin{equation}
\text{if}\,\,0\le x\le1,\,\text{then}\,\,\bar{x}\equiv 1-x.
\end{equation}

Eq. (5.4) is a bit involved and, even worse, contains the unknown
parameters $p_{Ai}$, $p_{Bi}$ implicitly.  But it can be simplified
by reducing its right hand side and thereby strengthening the
requirement on $C'$ a bit.  This is carried out in Appendix C, with
the conclusion that, if $\alpha\ge\beta$, then
\begin{equation}
\begin{split}
P_A+P_B\ge1:C'\le2(\gamma\bar{\gamma})^{1/2}((\bar{P}_B/\bar{P}_A)^2-1)\\(P_A-P_B)(P_B/P_A)/
\left[\left(\frac{P_A}{P_B}\frac{\bar{P}_B}{\bar{P}_A}\right)^2-1\right]\\
P_A+P_B\le1:C'\le2(\gamma\bar{\gamma})^{1/2}((P_A/P_B)^2-1)\\(P_A-P_B)(\bar{P}_A/\bar{P}_B)/
\left[\left(\frac{P_A}{P_B}\frac{\bar{P}_B}{\bar{P}_A}\right)^2-1\right]\\
\text{where}\,\,\gamma=N_A/N
\end{split}
\end{equation}
are sufficient to carry out the apparent reversal of ranking of $A$
and $B$.

Let us take a simple example that has been previously quoted [4]
[8]. We will paraphrase it and use rounded off data.  Hospitals $A$
and $B$ specialize in treating a certain deadly disease.  $N_A=1000$
patients are treated at $A$ and $N_B=1000$ at $B$.  Of these,
$S_A=900$ recover, while $S_B=800$ recover, so that $P_A=.9$,
$P_B=.8$ and Hospital $A$ is apparently the place to go.  In fact,
one computes $C_{AB}=.05$, so that this conclusion is supported at
the $.05\times(2000)^{1/2}=2.24$ standard deviation level.  Detailed
investigation shows that matters are not so simple.  Some patients
enter in otherwise good shape, others in poor shape.  Of the former,
$N_{A1}=900$ enter hospital $A$, and 870 recover; of the latter,
$N_{A2}=100$ enter and 30 recover, so $P_{A1}=.967$, $P_{A2}=.3$.

\begin{table}[h]
\caption{Simplified Hospital Recovery Data}\label{Ta:2}
\begin{center}
\begin{tabular}{l c c }
&Good Shape &Poor Shape\\
Admissions to Hospital A&900 &100\\
Recovered in Hospital A&870 &30\\
Admissions to Hospital B&600 &400\\
Recovered in Hospital B &590 &210\\
\multicolumn{3}{r}{Total Recovered/Admissions in A: 900/1000=.9}\\
\multicolumn{3}{r}{Total Recovered/Admissions in B: 800/1000=.8}\\
\end{tabular}
\end{center}
\end{table}

On the other hand, $N_{B1}=600$ enter Hospital $B$ in good shape and
$S_{B1}=590$ recover, whereas $N_{B2}=400$, $S_{B2}=210$. Thus,
$P_{B1}=.983$, $P_{B2}=.55$.  We see that by not mixing the two
classes of patients, Hospital $B$ is superior for each class -- at
levels $C'_1=.038$ (1.7 standard deviations) and $C'_2=.176$ (7.9
standard deviations).  Simpson, or inverse Simpson, depending upon
one's point of view, is certainly exemplified.

Of course, the criteria as to which patients entered in good shape,
which in poor shape, are a bit fuzzy.  Given the aggregate data, the
decomposition into the two classes could, as we have seen, been
planned with the intention of most convincingly asserting the
opposite of the conclusion from the aggregate data.  If this had
been done according to the prescription of (5.7), then with the same
input data, we would have found $\alpha=.935$, $\beta=.738$ (not far
from the $\alpha=.9$, $\beta=.6$ corresponding to the additional
data presented) and concluded with the superiority of Hospital $B$
at a confidence level corresponding to $C'\le.107$ or 4.79 standard
deviations for each class of patients.

\section{Concluding Remarks}

The Simpson paradox, one of the simplest examples of the common
misuse of statistics (think meta-analysis?) has received increasing
attention, since the consequences of its use -- or misuse -- can be
quite severe (as well as profitable).  In the classical Simpson
Paradox, the only question is whether or not to combine data from
different sources (and trying to justify the decision to combine).
What we have seen here is that the inverse Simpson paradox, even in
its most ``sophisticated'' version in which mean differences are
weighted by appropriate standard deviations, is nearly universally
applicable. This can be an effective analytical tool, but can
equally well be an effective technique for distorting statistical
data.

\appendix
\section{Evaluation of (4.9)}
Choosing Bayes with a uniform prior on $p_A,p_B$ space, (4.9)
becomes
\begin{equation}
\begin{split}
\sideset{}{}\iint_{\substack{1\ge p_A\ge p_B\ge 0}}
Pr\left(S_A,S_B|p_A,p_B,N_A,N_B\right)dp_A\,dp_B\left/\right.\\
\sideset{}{}\iint_{\substack{1\ge p_A, p_B\ge0}}
Pr\left(S_A,S_B|p_A,p_B,N_A,N_B\right)dp_A\,dp_B\\
=\sideset{}{}\iint_{\substack{1\ge p_A\ge p_B\ge0}}
p_A^{S_A}\,p_B^{S_B}\,q_A^{F_A}\,q_B^{F_B}\,dp_A\,dp_B\left/\right.\\
\sideset{}{}\iint_{\substack{1\ge p_A, p_B\ge 0}}
p_A^{S_A}\,p_B^{S_B}\,q_A^{F_A}\,q_B^{F_B}\,dp_A\,dp_B\\
=\int^1_0\left(\int_0^{p_A}p_B^{S_B}q_B^{F_B}dp_B\right)
p_A^{S_A}q_A^{F_A}dp_A\left/\right.\\
\int^1_0\!\int
p_B^{S_B}\,q_B^{F_B}\,p_A^{S_A}\,q_A^{F_A}\,dp_B\,dp_A\\
=\int^1_0 B_{p_A}\left(S_A+1, F_B+1\right)
p_A^{S_A}\,q_A^{F_A}\,dp_A/\\B\left(S_B+1, F_B+1\right)B
\left(S_A+1,F_A+1\right).
\end{split}
\end{equation}
Applying the known expansion of the incomplete Beta function [2],
this reduces after a little algebra to
\begin{equation}
\begin{aligned}
Pr(p_A&\ge p_B|S_A,S_B,N_A,N_B)\\
&=\sum^{F_A+F_B}_{j=0}\left(\begin{smallmatrix}{S_A+S_B+1+j}\\
{S_A}\end{smallmatrix}\right)\left(\begin{smallmatrix}{F_A+F_B-j}\\
{F_A}\end{smallmatrix}\right)\\&/\left(\begin{smallmatrix}{N_A+N_B+2}\\
{N_A+1}\end{smallmatrix}\right),
\end{aligned}
\end{equation}
or introducing $S=S_A+S_B$, $N=N_A+N_B$ for notational convenience,
\begin{equation}
\begin{aligned}
Pr(p_A&\ge p_B|S_A,S_B,N_A,N_B)\\
&=\sum^F_{j=0}\left(\begin{smallmatrix}{S+1+j}\\
{S_A}\end{smallmatrix}\right)\left(\begin{smallmatrix}{F-j}\\
{F_A}\end{smallmatrix}\right)/\left(\begin{smallmatrix}{N+2}\\
{N_A+1}\end{smallmatrix}\right)
\end{aligned}
\end{equation}

But we will go to the large sample limit defined by fixed
\begin{equation}
\begin{aligned}
s=N_A^{-1/2}\left(S_A-\frac{1}{2}N_A\right),
\qquad &\gamma=N_A/N,\\
s'=N_B^{-1/2}\left(S_B-\frac{1}{2}N_B\right)\qquad &1-\gamma\equiv
N_B/N
\end{aligned}
\end{equation}
as $N\to\infty$.  We could proceed precisely as in (4.5 -- 4.8), but
if we imagine a large sample limit from the outset, the derivation
is brief and standard.  Consider drug $A$.  A uniform prior for
$p_A$ is given by the beta distribution
\begin{equation}
\begin{aligned}
f(p_A)&=b(1,1;p_A)\\
\text{where}\,b(m,n; p_A)&=p_A^{m-1}q_A^{n-1}/B(m,n)\\
B(m_1n)&=m-1!n-1!/m+n-1!\\
\end{aligned}
\end{equation}
which, after$S_A$ successes in $N_A$ trials creates the posterior
distribution
\begin{equation}
b(1+S_A, 1+N_A-S_A; p_A).
\end{equation}
Drug B works the same way.  It follows that
\begin{equation}
\begin{aligned}
E(p_A-p_B)&=\frac{S_A+1}{N_A+1}-\frac{S_B+1}{N_B+1}\\
\text{Var}(p_A-p_B)&=\frac{(S_A+1)(N_A+1-S_A)}{(N_A+1)^2(N_A+2)}\\
&+\frac{(S_B+1)(N_B+1-S_B)}{(N_B+1)^2(N_B+2)},
\end{aligned}
\end{equation}
and so by the central limit theorem for large $N_A$, $N_B$,
\begin{equation}
\begin{aligned}
Pr(p_A\ge
p_B)=\phi\left(\frac{S_A}{N_A}-\frac{S_B}{N_B}/\left(\frac{S_A(N_A-S_A)}{N_A^3}\right.\right.\\
\left.\left.+\frac{S_B(N_B-S_B)}{N_B^3}\right)^{1/2}\right)\\
\text{where}\, \phi(x)=\frac{1}{\sqrt{2\pi}}\int_{-\infty}^x
e^{-\frac{1}{2}y^2}dy.
\end{aligned}
\end{equation}

\section{Restrictions on $C'$}
Eq. (5.3) itself imposes two conditions.  Aside from the crucial
$0\le P_{A1},P_{A2},\\ P_{B1}, P_{B2}\le1$, there are just two more
due to the composition conditions that
$S_{A1}+S_{A2}=S_A,\,N_{A1}+N_{A2}=N_A,\,S_{B1}+S_{B2}=S_B,\,N_{B1}+N_{B2}=N_B$.
We reintroduce the notation of Section 2:
\begin{equation}
\begin{aligned}
N_{A1}=\alpha N_A,& \qquad N_{B1}=\beta N_B
\end{aligned}
\end{equation}
and hereafter uniformly adopt the notation that
\begin{equation}
\begin{aligned}
\text{if}\,\,0\le x\le 1,\qquad\text{then}\,\,\bar{x}\equiv 1-x.
\end{aligned}
\end{equation}
Thus $S_{A1}+S_{A2}=S_A$ implies $P_{A1} N_{A1}+P_{A2}
N_{A2}=P_AN_A$, or
\begin{equation}
\alpha P_{A1}+\bar{\alpha}P_{A2}=P_A
\end{equation}
and similarly
\begin{equation}
\beta P_{B1}+\bar{\beta}P_{B2}=P_B.
\end{equation}
We also append (5.3) in the form
\begin{equation}
\begin{aligned}
P_{B1}-P_{A1}&=C'\sigma_1\\
P_{B2}-P_{A2}&=C'\sigma_2,
\end{aligned}
\end{equation}
and solve (B.3), (B.4), (B.5) to yield
\begin{equation}
\begin{aligned}
P_{A1}&=K_1+\frac{\bar{\alpha}}{\alpha-\beta}\,C'\sigma_B,&
P_{A2}&=K_2-\frac{\alpha}{\alpha-\beta}\,C'\sigma_B,\\
P_{B1}&=K_1+\frac{\bar{\beta}}{\alpha-\beta}\,C'\sigma_\alpha,&
P_{B2}&=K_2-\frac{\beta}{\alpha-\beta}\,C'\sigma_\alpha\\
\end{aligned}
\end{equation}
where
\begin{equation}
\begin{aligned}
K_1&=(\bar{\beta}P_A-\bar{\alpha}P_B)/(\alpha-\beta),&\,\,
\sigma_\alpha&=\alpha\sigma_1+\bar{\alpha}\sigma_2,\\
K_2&=(\alpha P_B-\beta P_A)/(\alpha-\beta), &\,\,
\sigma_\beta&=\beta\sigma_1+\bar{\beta}\sigma_2.
\end{aligned}
\end{equation}

Eqs. (B.6), (B.7) are realizable if the requirements $0\le P_{A1},
P_{A2}, P_{B1}, P_{B2}\le1$ are satisfied.  Since we are asserting,
without loss of generality, that $p_A\ge p_B$, we of course have the
condition
\begin{equation}
P_A\ge P_B,\bar{P}_B\ge\bar{P}_A.
\end{equation}
There are then two cases to consider.  If $\alpha\ge\beta$, it is
easily seen that $K_1\ge0$, $K_2\le1$, so that $P_{A1}, P_{B1}\ge
0$, $P_{A2}, P_{B2}\le1$ are already satisfied.  The remaining four
conditions $P_{A1}, P_{B1}\le 1$, $P_{A2}, P_{B2}\ge0$ can then be
gathered together as
\begin{equation}
\begin{aligned}
\text{if}\,\alpha\ge\beta\,\text{then}\qquad\qquad\qquad\qquad\qquad\qquad\qquad\\
C'\le\min\left(\frac{(\alpha-\beta)(1-K_1)}{\bar{\alpha}\sigma_\beta},\frac{(\alpha-\beta)(1-K_1)}{\bar{\beta}\sigma_\alpha},\right.\\
\left.\frac{(\alpha-\beta)K_2}{\alpha\sigma_\beta},\frac{(\alpha-\beta)K_2}{\beta\sigma_\alpha}\right),
\end{aligned}
\end{equation}
or, inserting (B.7),
\begin{equation}
\begin{aligned}
C'=\min\left(\frac{\bar{\beta}\bar{P_A}-\bar{\alpha}\bar{P_B}}{\bar{\alpha}\sigma_\beta},\frac{\bar{\beta}
\bar{P_A}-\bar{\alpha}\bar{P_B}}{\bar{\beta}\sigma_\alpha},\right.\\
\left.\frac{\alpha P_B-\beta P_A}{\alpha\sigma_\beta},\frac{\alpha
P_B-\beta P_A}{\beta\sigma_\alpha}\right).
\end{aligned}
\end{equation}
Similarly,
\begin{equation}
\begin{aligned}
\text{if}\,\alpha\le \beta\,\text{then}\qquad\qquad\qquad\qquad\qquad\qquad\\
C'\le\min\left(\frac{\bar{\alpha}P_B-\bar{\beta}P_A}{\bar{\alpha}\sigma_\beta},\frac{\bar{\alpha}P_B-\bar{\beta}P_A}
{\bar{\beta}\sigma_\alpha},\right.\\
\left.\frac{\beta\bar{P_A}-\alpha\bar{P_B}}{\alpha\sigma_\beta},\frac{\beta\bar{P_A}-\alpha\bar{P_B}}
{\beta\sigma_\alpha}\right)
\end{aligned}
\end{equation}
Since we require $C'\ge0$, immediate consequences are that
\begin{equation}
\begin{aligned}
&\text{if}\,\,\alpha\ge\beta,\text{then}\,\,
\frac{\alpha}{\beta}\ge\frac{P_A}{P_B}\ge 1,\qquad
\frac{\bar{\beta}}{\bar{\alpha}}\ge\frac{\bar{P_B}}{\bar{P_A}}\ge 1\\
&\text{if}\,\,\alpha\le
\beta,\,\,\frac{\bar{\alpha}}{\bar{\beta}}\ge\frac{P_A}{P_B}\ge1,\,\,\frac{\beta}{\alpha}\ge
\frac{\bar{P_B}}{\bar{P_A}}\ge 1
\end{aligned}
\end{equation}
must hold.

\section{Simplification of (5.4)}
The major step is the observation, from (5.2) that
\begin{equation}
\sigma^2_i\le\frac{N}{4}\left(\frac{1}{N_{Ai}}+\frac{1}{N_{Bi}}\right),
\end{equation}
so that
\begin{equation}
\begin{aligned}
\sigma^2_1&\le\frac{N}{4}\left(\frac{1}{\alpha
N_A}+\frac{1}{\beta N_B}\right)\\
\sigma^2_2&\le\frac{N}{4}\left(\frac{1}{\bar{\alpha}N_A}+\frac{1}{\bar{\beta}N_B}\right)
\end{aligned}
\end{equation}
Hence,
\begin{equation}
\begin{aligned}
\text{if}\,\alpha\ge\beta,\qquad\sigma^2_1&\le\frac{1}{4\beta}\left(\frac{N}{N_A}+\frac{N}{N_B}\right)\\
\qquad\sigma^2_2&\le\frac{1}{4\bar{\alpha}}\left(\frac{N}{N_A}+\frac{N}{N_B}\right),
\end{aligned}
\end{equation}
yielding
\begin{equation}
\begin{aligned}
\begin{matrix}\sigma_2\\
\sigma_B\end{matrix}\}\le\max(\sigma_1,\sigma_2)\le\frac{1}{2}\left(\frac{N}{N_A}+\frac{N}{N_B}\right)^{1/2}\\
\max\left(\frac{1}{\beta^{1/2}},\frac{1}{\bar{\alpha}^{1/2}}\right).
\end{aligned}
\end{equation}
Setting $N_A/N=\gamma$, condition (5.4) can therefore be
strengthened to
\begin{equation}
\begin{aligned}
\alpha\ge \beta:C'\le2(\gamma\bar{\gamma})^{1/2}\min(\beta,\bar{\alpha})^{1/2}\\
\min\left[\frac{1}{\bar{\beta}}(\bar{\beta}\bar{P_A})
-\bar{\alpha}\bar{P_B}),\frac{1}{\alpha}(\alpha P_B-\beta
P_A)\right].
\end{aligned}
\end{equation}
And in the same way, we obtain the strengthened
\begin{equation}
\begin{aligned}
\alpha\le B:C'\le 2(\gamma\bar{\gamma})^{1/2}\min(\alpha,
\bar{\beta})^{1/2}\\
\min\left[\frac{1}{\bar{\alpha}}(\bar{\alpha}P_B-\bar{\beta}P_A),
\frac{1}{\beta}(\beta\bar{P}_A-\alpha\bar{P}_B)\right].
\end{aligned}
\end{equation}

Eqs. (C.5) and (C.6) are valid for all $\alpha,\beta$, and we may
indeed find the largest feasible range for $C'$ by maximizing their
right hand sides over $\alpha$ and $\beta$.  Again, to reduce
complexity, let us take the special case in which:
\begin{equation}
\alpha\ge\beta:\bar{\alpha}/\bar{\beta}=(\bar{P_A}/\bar{P_B})^2,
\beta/\alpha=(P_B/P_A)^2
\end{equation}
so that
\begin{equation}
\begin{aligned}
\alpha=[(P_A/P_B)^2(\bar{P_B}/\bar{P_A})^2-(P_A/P_B)^2]/\\
[(P_A/P_B)^2(\bar{P_B}/\bar{P_A})^2-1]\\
\beta=[(\bar{P_B}/\bar{P_A})^2-1]/[(P_A/P_B)^2(\bar{P_B}/\bar{P_A})^2-1]
\end{aligned}
\end{equation}
converting (C.5) and (C.6) to
\begin{equation}
\begin{split}
\alpha\ge\beta:C'\le
2(\gamma\bar{\gamma})^{1/2}/[(P_A/P_B)^2(\bar{P_B}/\bar{P_A})^2-1]\\
\min[(\bar{P_B}/\bar{P_A})^2-1,(P_A/P_B)^2-1]\\
\cdot\min(\bar{P_A}-\bar{P_A}^2/\bar{P_B}, P_B-P_B^2/P_A).
\end{split}
\end{equation}
But
$(\bar{P_A}-\bar{P_A}^2/\bar{P_B})-(P_B-P_B^2/P_A)=(1-P_A-P_B)(P_A+P_B)^2/P_A\bar{P_B}$
and
$((\bar{P_B}/\bar{P_A})^2-1)-((P_A/P_B)^2-1)=(P_A+P_B-1)\frac{P_A-P_B}{P_A\bar{P_B}}\left(\frac{P_A}{P_B}+
\frac{\bar{P_B}}{\bar{P_A}}\right)$, so it follows that in the
$\alpha\ge\beta$ case,
\begin{equation}
\begin{aligned}
P_A+P_B\ge1:C'\le2(\gamma\bar{\gamma})^{1/2}\left(\left(\frac{\bar{P_B}}{\bar{P_A}}\right)^2-1\right)\\
(P_A-P_B)\frac{P_B}{P_A}/\left[\left(\frac{P_A}{P_B}\frac{\bar{P_B}}{\bar{P_A}}\right)^2-1\right]\\
P_A+P_B\le1:C'\le2(\gamma\bar{\gamma})^{1/2}\left(\left(\frac{P_A}{P_B}\right)^2-1\right)\\
(P_A-P_B)\frac{\bar{P_A}}{\bar{P_B}}/\left[\left(\frac{P_A}{P_B}\frac{\bar{P_B}}{\bar{P_A}}\right)^2-1\right]
\end{aligned}
\end{equation}
are sufficient to carry out the apparent reversal of ranking of $A$
and $B$.  The decomposition corresponding to the choice $\alpha\le
\beta$ can of course be similarly specialized.

\end{document}